\documentclass[fleqn,10pt]{wlscirep}
\usepackage[utf8]{inputenc}
\usepackage[T1]{fontenc}
\usepackage{amsfonts, graphicx, physics, hyperref, empheq, lipsum, xcolor}
\usepackage{academicons}
\usepackage[normalem]{ulem}
\hypersetup{
	colorlinks=true,
	linkcolor=blue,
	filecolor=magenta,      
	urlcolor=cyan,
	citecolor=blue, 
}
\newcommand{\add}[1]{#1}

\title{Detecting Initial Correlations via Correlated Spectroscopy in Hybrid Quantum Systems}

\author[1,2,*]{Parth Jatakia}
\author[1,3]{Sai Vinjanampathy}
\author[4]{Kasturi Saha}
\affil[1]{Department of Physics, Indian Institute of Technology Bombay, Mumbai , India}
\affil[2]{Department of Electrical Engineering, Princeton University, Princeton, New Jersey 08540, USA}
\affil[3]{Centre for Quantum Technologies, National University of Singapore, Singapore}
\affil[4]{Solid State Device Group, Department of Electrical Engineering, Indian Institute of Technology Bombay, Mumbai , India}
\affil[*]{pjatakia@princeton.edu}

\begin{abstract}
Generic mesoscopic quantum systems that interact with their environment tend to display appreciable correlations with environment that often play an important role in the physical properties of the system. However, the experimental methods needed to characterize such systems either ignore the role of initial correlations or scale unfavourably with system dimensions. Here, we present a technique that is agnostic to system-environment correlations and can be potentially implemented experimentally. Under a specific set of constraints, we demonstrate the ability to detect and measure specific correlations. We apply the technique on a hybrid quantum system of Nitrogen Vacancy Centers (NV) coupled to an optical cavity with initial correlations. We extract the interaction strength and effective number of interacting NVs from the initial correlations using our technique.
\end{abstract}
\begin{document}

\flushbottom
\maketitle

\thispagestyle{empty}

\section{Introduction}
\label{intro}
Quantifying complex dynamics of a correlated system and environment is a necessary step for realizing practical quantum technologies. Traditionally, majority of the methods describing the dynamical evolution of an open-quantum system assume factorization approximation \cite{li2018concepts}, which states that the initial system-environmental correlation is negligible. The advantage in assuming absence of correlations between system and environment is the guarantee that the evolution of the reduced system is given by completely positive trace preserving maps (CPTP) between the reduced system states \cite{BRE02}. A typical dynamical equation that appears for describing complex quantum systems derived under factorization approximation alongside Markov approximation and rotating wave approximation is the GKLS (Gorini-Kossakowski-Lindblad-Sudarshan) master equation \cite{Lindblad1976}. However, these descriptions fail in the presence of initial correlations, which are present in strongly correlated nanoscale system-environment dynamics \cite{Ishizaki2010,GonzlezTudela2017,Pollock2018}.

\add{For such initially correlated system-environment dynamics}, the reduced dynamical map describing the system evolution is described by ``not completely positive" (NCP) maps. NCP-maps capture the effect of initial correlations on the subsequent dynamics, since in the presence of correlations, the marginal states and the correlation matrix are not independent. Such NCP-maps have been discussed in the literature from a mathematical point of view \cite {PhysRevA.97.032127,shaji2005s,jordan2004dynamics}, and several theoretical results have been established relating to the violation of laws of physics as a consequence of NCP-maps \cite{buscemi2014complete,vinjanampathy2016correlations}. Process tensor tomography does not assume CP maps and is often used to characterize such maps. While state tomography of a $d\times d$ density matrix scales as $\mathcal{O}(d^2)$,  process tomography scales as $\mathcal{O}(d^4)$ and process tensor tomography scales as $\mathcal{O}(d^6)$ \cite{Ringbauer2015}, making it impractical to characterize even modest systems. Thus experimental efforts till date for characterization have been majorly capable of only witnessing the presence of NCP maps by detecting initial correlations. Most of these techniques \cite{gessner2014local,smirne2011experimental,li2011experimentally} rely on measurement of the change in trace distance in time between two states which initially have same system marginal states. Another method \cite{yu2019experimental} to witness initial correlation is by measuring the conditional past-future correlations. 

The fact that characterizing NCP dynamics is cumbersome has direct consequences on several forms of spectroscopic techniques used. The experimental inconvenience has forced these techniques to be based on local master equations (like GKLS) that are based on CP maps formalism which in turn rely on Hilbert space dimensionality of the system and the Born-Markov assumption. Hence, it is essential to develop spectroscopic techniques that neither rely \add{strongly on number of dimensions of the Hilbert space of the system} nor on whether these degrees of freedom are interacting with an environment.

In this work, we propose a new method we call Prepare-Probe-Spectroscopy (PrePSy). Using PrePSy, we show selective detection of particular system-environment correlations and furthermore quantify them under some assumptions. Our spectroscopic method extracts information about the system-environmental initial correlations by utilizing the role measurements play on correlated systems. Such measurement based methods were previously used to characterize NCP maps \cite {modi2012operational}, modify information theoretic bounds \cite {vinjanampathy2015entropy} and witness initial correlations \cite {nahar2018preparations}. The spectroscopic technique applied in PrePSy additionally facilitates deriving information related to system-environment dynamics embedded in the initial correlations. This functionality of PrePSy is demonstrated using a spin-photon hybrid quantum system-environment represented by color centers present in diamonds \cite {Aharonovich2011} or SiC \cite {Calusine2014,Sipahigil2014} placed in an optical resonator. By implementing PrePSy on this setup, information such as the photon-mediated interaction strength between two color centers and decoherence led effects are derived. Such hybrid QED setups have been used to implement quantum information processing applications \cite {Kok2007,Kimble2008}, quantum key distribution \cite {Lo2014,Takemoto2015} and entanglement distribution \cite {Hensen2015,Stockill2017,Delteil2015}. Among various attempts for constructing such solid-state cavity QED systems, NVs coupled to a photonic crystal cavity \cite {Englund2010,Barth2009}, microsphere resonator \cite {Park2006,Schietinger2008}, or micro-toroids \cite {Gregor2009,Barclay2009} have emerged as a promising candidate. The prevalence of color centers for quantum information applications is because of their spin degree of freedom, which has advantages such as a long spin decoherence time at room temperature \cite {Balasubramanian2009,Widmann2014,Maurer2012} and convenient optical readout of spin state \cite {Fuchs2009} in addition to the photonic degree of freedom.

The organization of the manuscript is as follows - in the section \ref{sec:2}, we describe the three steps - Prepare, Evolve, and Probe, that constitute PrePSy. In section \ref{sec:3}, the PrePSy technique is demonstrated using a toy model to detect and measure initial correlations. In section \ref{sec:4}, the capability of PrePSy to decode information related to dynamics of system-environment from the initial correlations is displayed for multiple NVs positioned within a photonic cavity signifying a spin-photon hybrid system. Information such as the coupling constant and effective number of NVs interacting given the decoherence is extracted.

\section{Prepare Probe Spectroscopy (PrePSy)}
\label{sec:2}
To study systems which are initially correlated with their environment, it is well known that operation of an entanglement breaking channel \cite{horodecki2003entanglement,modi2012operational,vinjanampathy2016correlations} on the system can be used to generate environmental states conditioned on the outcome of the system states. In addition to entanglement breaking channel, we use a two dimensional spectroscopic technique to probe dynamics of the system interacting with a conditioned state of the environment. We demonstrate how the two techniques in tandem can detect and measure initial or intermediate correlations in quantum systems alongside detecting the hidden Hilbert space dimensionality of the environment.

\subsection{Step 1 : Conditional Preparation}

The first step entails preparing the system and environment state. In this step, a projective measurement  is performed on the system which acts as an entanglement breaking channel. This is followed by a unitary transformation of the resultant system marginal state to a standard state.

For a mathematical description of this step, consider a general initial state of the system-environment written as 
\begin{equation}
    R=\rho\otimes\tau + \chi,
    \label{eqn:postcp}
\end{equation}
where $R$ is the total state of the system-environment, $\rho$ and $\tau$ are the marginal state of the system and environment respectively, and $\chi$ is the correlation matrix. The projective measurement $\mathcal{E}_m$ projects the system state onto $\ket{m}$, which leads to
\begin{equation}
    \mathcal{E}_m[R]=\ketbra{m}{m}\otimes\sigma^{(m)}.
\end{equation}
Here $\sigma^{(m)} = \tau + \text{Tr}_s(\ketbra{m}{m}\otimes \mathbb{I}\cdot\chi) $ is the post measurement state of the environment where $\text{Tr}_s(\cdot)$ is the partial trace over the system dimensions. Thus in presence of initial system-environment correlations, projective measurement performed on the system generates environment state conditioned on the measurement. The choice of projective measurement operator is crucial as it selectively allows parts of the initial correlation matrix $\chi$, that are along the projective operator as shown in section 1 of Supplementary.

\add{Next, the system state is prepared by a unitary transformation to a fiducial state $\ket{0}$.} The total state after preparation is as follows,
\begin{equation}
    \mathcal{U}_m\circ \mathcal{E}_m[R]=\ketbra{0}{0}\otimes \sigma^{(m)},
    \label{eqn:rho0}
\end{equation}
where $\mathcal{U}_m$ transforms system state $\ket{m}$ to $\ket{0}$.  \add{Hence, after preparation any difference between two states which were projected onto different outcomes ($\ket{m}$,$\ket{n}$), is only between the environmental states ($\sigma^{(m)}$ and $\sigma^{(n)}$).} This difference between the environmental marginal states is caused only due to the presence of initial correlations. We perform step 1 multiple times with variations in the states chosen for projective measurement so that pairwise differencing as shown later in Eqn. \ref{eqn:differencing} can be applied. We choose orthogonal states for the set of variations in projective measurements as it facilitates generation of independent data sets. The number of maximum possible iterations scales favourably with the dimensions of the system only.

\subsection{Step 2 : Two Dimensional Spectroscopy}
The next step is inspired from 2D Phase coherent spectroscopy which is routinely deployed for Nuclear Magnetic Resonance (NMR) \cite{levitt_2015,Aue1976}. An elementary example of 2D spectroscopy is shown in Figure \ref{fig:PrePSy}.a. The rotation caused by the first control-pulse is for preparation of spins. The delay $t_1$ and $t_2$ is individually varied and separated by a similar rotation generating control pulse. Consequently the signal is recorded by measuring an appropriate property for e.g. magnetization is measured for nuclear spin ensemble. The signal will be a function of $t_1$ and $t_2$, and on Fourier transformation will generate a 2D spectrum common to spectroscopy. Though one-dimensional spectroscopy suffices the need for characterizing initial correlations, we choose two-dimensional spectroscopy. The reason is that with two-dimensional spectroscopy in addition to characterizing correlations we can also predict system-environment dynamics based on the correlations measured. Thus the added benefit of 2D spectroscopy is measuring coherent population transfer channels i.e., valid transitions, in a multipartite system required for calculating the system-environment dynamics.

Step 1 - ``Conditional Preparation"  is integrated with the two-dimensional spectroscopy by substituting the step 1 in place of the first pulse of the pulse sequence for two dimensional spectroscopy as shown in Fig \ref{fig:PrePSy}.a. The substitution generates a procedure as shown in the first and second part of Fig. \ref{fig:PrePSy}.b. Thus after the conditional preparation, the system and environment evolve for time $t_1$ followed by a $\pi/2$ pulse generated by suitable rotation operator ($\hat{A}$). We use the NMR definition of $\pi/2$ pulse and loosely extend it to any system Hilbert space to denote a $\pi/2$ rotation in a plane defined by the operator $\hat{A}$ acting on the prepared state. Here, $\frac{\pi}{2}$ pulse implies traversing half the angle between two states, preferably orthonormal to each other. \add{The angle subtended between them is $\theta = 2*\text{cos}^{-1}(\left|\braket{\psi}{\gamma}\right|)$ where $\ket{\psi}$ and $\ket{\gamma}$ are normalized states.}. This is then followed by free evolution for time $t_2$. 

\subsection{Step 3: Measurement}
\begin{figure*}[htb]
\includegraphics[width=1\textwidth]{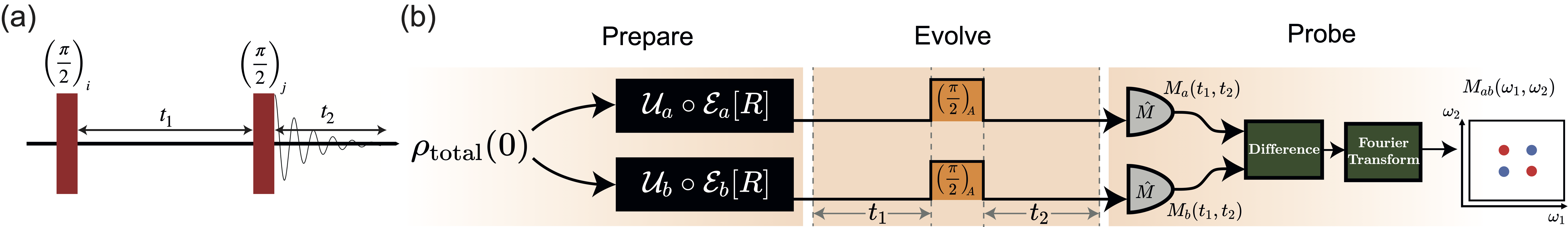}
\caption{\label{fig:PrePSy}(a) A generic pulse sequence for two-dimensional spectroscopy. $i$ and $j$ denotes the direction of the rotation caused by the pulse. The most commonly used setting is $i=j=x$ called the cosine version of COSY. (b) Schematic of the Preparation Probe Spectroscopy (PrePSy). The first part is the preparation part where conditional measurement is performed. The second part is the evolve part which is a time delayed pulse and time delayed measurement. Finally difference of the measured signals is taken followed by a Fourier transform. }
\end{figure*}

The final step is probing/measuring the system with a suitably chosen system observable. This system observable is kept constant for all the runs of variations of the projective measurement performed in step 1.  From the set of these variations, pairs are created for their respective measured signal to be differenced as follows
\begin{equation}
\label{eqn:differencing}
    M_{ij}(t_1,t_2) = M_i(t_1,t_2) - M_j(t_1,t_2),
\end{equation}
where $M_i(t_1,t_2)$ is the measured signal where in step 1 the projective measurement was performed along $\ket{i}$. Presence of initial correlation leads to difference in the environmental marginal states after step 1. This type of difference is now also seen in the system marginal states due to spectroscopy performed in step 2. Thus pairwise differencing at the end implies that effectively we are observing the result of spectroscopy performed on only the terms representing the presence of initial correlations. Each of these pairs is then Fourier transformed to generate the spectrum. The maximum number of pairs possible is $^dC_2$, however all pairs are not required. \add{Thus PrePSy scales as $\mathcal{O}(d^2)$ with the system dimensions.} Figure \ref{fig:PrePSy}.b shows the complete procedure for Prepare Probe Spectroscopy (PrePSy).

For zero initial correlations, the total state after step 1 (conditional preparation) will be the same for any projective measurement outcome. Thus the total state will also be the same after step 2 (spectroscopy) for any projection operator chosen in step 1 (conditional preparation). This is because the same process is performed for all states. Hence for any two pairs of different projection operators chosen in step 1 (conditional preparation), the pairwise difference in step 3 (measurement) will give zero if there is no initial correlation. \add{Thus if PrePSy detects some signal, it implies the presence of correlations.} An appropriate choice of the projective measurement, rotation operator of the $\pi/2$ pulse and final measurement operators will ensure that the initial correlations are always almost captured by PrePSy as shown in section 1 \& 2 of Supplementary. \add{This will reduce the need to measure all $^dC_2$ pairs.}

\add{The scaling of PrePSy as $\mathcal{O}(d^2)$ is better than process tensor tomography ($\mathcal{O}(d^4)$). This independence from the environmental dimension provides a huge benefit experimentally. Moreover if the entities in the environment interacting with the system increase then on an average the correlations with the system will mostly decrease. This effect is due to the fact that a small number of environment spins approximates a thermal bath under generic interactions. For typical Hamilonians the correlations always decrease with the increase in dimension of the environment generating thermalization \cite{dubey2012approach}. This renders PrePSy and other system-environment characterizing tools irrelevant in this regime of large environment dimensions.}

\add{PrePSy scales not only with the system dimensions but also with sampling being done in $t_1$ and $t_2$. For a sampling size "$S$" for $t_1$ $\&$ $t_2$ and "$A$" number of averages per sample, there needs to be $(A \times S)^2$ measurements. Hence the mean number of runs required such that all the data points for a particular choice of preparation is $({1}/{P(a)})^{(AS)^2}$ where $P(a)$ is the probability of state “$a$”. Thus the mean number of runs required for a particular choice of preparation depends directly on the sampling size and indirectly on the system dimensions ($P(a)$ depends on system dimensions). Thus PrePSy scales exponentially with the sampling size.}

\section{Toy Model}
\label{sec:3}
In this section, we devise a toy model of a system and environment with initial correlation and characterize it using PrePSy. A generic Hamiltonian for the system and the environment can be written as a sum of two parts. The first part is a diagonal Hamiltonian that defines the joint energy levels of the system and environment. The other part is an interaction Hamiltonian, which may be off-diagonal to promote transitions between different levels. We can write a generic density matrix ($R$) of such a system-environment in the Fano representation \cite{fano1957description} as
\begin{equation}
    R=\frac{1}{dD}(I+\vec{r}.\vec{M}\otimes I+I\otimes \vec{s}.\vec{N}+\sum_{ij}M_iT_{ij}N_j),
\end{equation}
where $\{M_{i=1\ldots d^2-1}\}$ and $\{N_{j=1\ldots D^2-1}\}$ are elements of a Hermitian traceless matrix basis, d(D) is the Hilbert space dimension of the system(environment) and $\vec{r}, \vec{s}$ and $T_{ij}$ are real coefficients such that R is a well-defined density operator. Using these guidelines, we construct our toy model.

Here we consider a toy model of a spin-qubit system coupled to a spin-qubit environment as an example and demonstrate the effectiveness of PrePSy by measuring initial correlations. Though our constructions are simple, \add{the derived conclusions hold true for higher dimensions.}

\subsection{Total Hamiltonian and Initial State}
A general two-qubit Hamiltonian can be written as
\begin{equation}
    H = \omega^{(s)}S_z^{(s)} +\omega^{(e)}S_z^{(e)} + \sum_{i\in \{x,y,z\}}\lambda^{(i)}S_i^{(s)}S_i^{(e)},
\end{equation}
where $S_i^{(s)}$ ($S_i^{(e)}$) is the spin$-1/2$ angular momentum of the system (environment), $\omega^{(s)}$ ($\omega^{(e)}$) is the energy difference between the two states of the system (environment) qubit and $\lambda^{(ij)}$ is the coupling between the two spins. 

For the initial state, the general density matrix of two qubits can be written as 
\begin{equation} \label{eqn:densitymatrix}
R = \frac{1}{4} \left(I + \vec{u}\vec{\sigma}\otimes I + I\otimes\vec{v}\vec{\sigma} + \sum_{j,k }T_{jk}\sigma^{s}_{j}\otimes\sigma^{e}_{k}\right),
\end{equation}
where $\vec{\sigma} = (\sigma_x,\sigma_y,\sigma_z)$ is a Pauli vector. For $R$ to be a well defined density matrix,  $\vec{u},\vec{v}\in \mathbb{R}^3$ and $T_{j,k} \in \mathbb{R}$ are appropriately chosen. Using singular value decomposition, matrix $T = \{T_{jk}\}$ can be written as $T = O^a\text{ diag}\{c_2,c_2,c_3\}O^b$ where $O^a$ and $O^b$ are orthogonal matrices. One can see that $T$ has local unitary equivalence to
\begin{equation}
    R = \frac{1}{4}\left(I + \vec{a}\vec{\sigma}\otimes I + I\otimes\vec{b}\vec{\sigma} + \sum_{j \in \{x,y,z\}}c_j\sigma_j\otimes\sigma_j \right),
\end{equation}
where $\vec{a},\vec{b},\vec{c}$ are real coefficients chosen such that $R$ is a valid density matrix. Thus a general two qubit state can always be reduced, up to local unitary equivalence, to a state in the above form, Since focus is on the correlations in the bipartite system, we set $\vec{a} = \vec{b} = 0$, thus considering only the maximally mixed marginals. The system-environment state is given as
\begin{equation}\label{rhomain}
    R = \frac{1}{4}\left(I + \sum_{j \in \{x,y,z\}}c_j\sigma_j\otimes\sigma_j\right).
\end{equation}

Here, $\{c_j\}$ are real parameters consistent with $R$ being a well-defined density operator. For the numerical example, arbitrarily but consistently chosen $c_j$ and $\{\omega^{(s)},\omega^{(e)},\lambda^{(ij)}\}$ are used. PrePSy requires minimum of two variations in the projection operator in Step 1 for pairwise differencing, and the number of orthogonal variations can go up to the size of the system. Here as the system is itself two dimensional, only two runs with projection operators $\ketbra{x}{x}$ and $\ketbra{-x}{-x}$ are performed. The $\pi/2$ pulse is chosen to be in a direction perpendicular to both the projective operators i.e. about the y-axis for an optimal rotation. The final measurement operator chosen is the population measurement of $\ket{x}$. In section 2 of Supplementary, we present a calculation to understand the relation between the various matrix elements discussed in producing the signal detected. The simulation performed assumes a closed unitary dynamics of two qubit.

\subsection{PrePSy on Toy Model}

\begin{figure}[htp!]
\centering
\includegraphics[width=0.95\columnwidth]{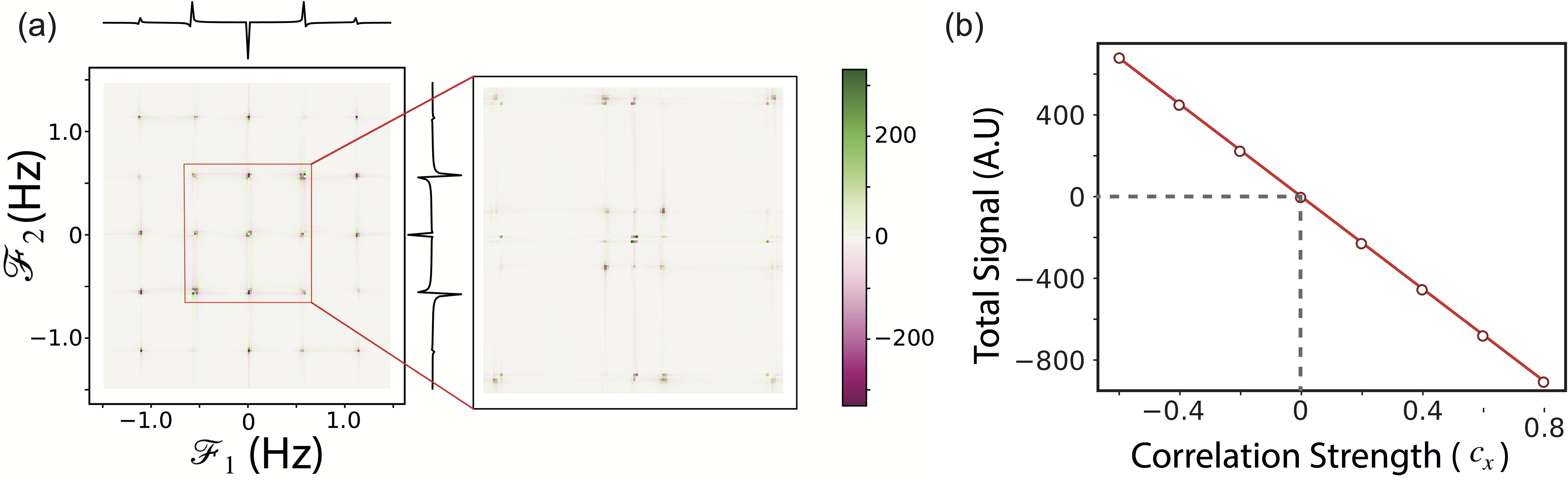}
\caption{\label{fig:xcorr_detect} (a) 2D data of PrePSy on the toy model with parameters as $c_x = -0.8$, \add{$c_y = 0$, $c_z = 0 $}, $\lambda^{(xx)} = 4 \text{ Hz}$,  $\lambda^{(yy)} = 3 \text{ Hz}$ and $\lambda^{(zz)} = 3.5 \text{ Hz}$. \add{A total of $150$ steps were simulated for variation in both $t_1$ and $t_2$.} Presence of peaks denotes existence of initial correlation. (b) Plot of the variation in total signal strength (sum of the 2D data) vs the initial correlation $c_x$.}
\end{figure}

For initial state with non-zero correlation, applying PrePSy gives the peaks shown in Figure \ref{fig:xcorr_detect}.a. Thus obtaining non-zero data in the form of peaks implies that it does detect correlation correctly as discussed previously.

The position of the peaks in the 2D diagram in Figure \ref{fig:xcorr_detect}.a describes the frequency corresponding to the energy gap of two states between which a population transfer occurred during $t_1$ and $t_2$. The peak positions depends on the interaction Hamiltonian which dictates the energy gap. The intensity of the peak represents the quantity of the population transferred and hence depends on the initial density matrix and thus the initial correlations. Upon measuring the variation of intensity of any peak  with the correlation strength, a linear dependency is revealed. This is shown in section 2 of Supplementary, where Eqn.7 in Supplementary implies that derivative of the signal measured with initial correlation is independent of the initial correlation proving the aforementioned statement. Linearity is not surprising because the master equation is linear in the density matrix which contains the correlation. Plot of total signal intensity (2D sum of the signal over all frequency) \textit{vs.} correlation is shown in Figure \ref{fig:xcorr_detect}.b.

If the equation of this line is known a priori, then by applying PrePSy and measuring the total intensity of the total signal generated by PrePSy, initial correlation can be calculated from the equation of the line. Thus to measure correlations the equation of line must be known.

To know the equation of a line two data points are required. Out of which first is trivial i.e. the line passes through (0, 0). One more data point, if acquired, should be sufficient. However, knowing the system-environment correlation a priori is nontrivial. If the Hamiltonian of the system - environment is entirely known then a simple method for calculating the other data point is using the Gibb's thermal state where initial correlation can be calculated. Performing PrePSy on the thermal state of the system and environment will give an additional data point. If the Hamiltonian is unknown, then the equation of the line defining the correlation for the measured signal cannot be estimated. In this case, for a small enough system-environment Hilbert space, brute force method can be used to learn Hamiltonian parameters from the PrePSy data.

As mentioned earlier one-dimensional spectroscopy suffices for characterizing initial correlations. The results of PrePSy, if 1D spectroscopy is performed on the toy model instead of 2D spectroscopy is shown in section 3 of Supplementary. Unlike 2D spectroscopy, for PrePSy with 1D spectroscopy, Hamiltonian parameters cannot be learned from the measured signal as the signal generated by a particular Hamiltonian is no longer unique. The loss of uniqueness is briefly discussed in the conclusion section.

\section{Application : Cavity QED with NVs}
\label{sec:4}
The measured initial correlations can provide insights to the system-environment dynamics. In this section, we show that the initial correlations measured by PrePSy can be used to estimate physical properties of the system-environment. We select multiple NVs coupled to a cavity because the quick characterization of the physical properties by PrePSy can help mitigate the experimental challenge in coupling them to cavities. Developing reliable interfaces between photons and the NVs \cite{wan2019large} are particularly tricky due to a significantly small fraction of fluorescence contributed by zero phonon line and low coupling strength of NVs interacting with cavity photon. Moreover, these cavities have dimensions in the nanoscale regime which generates low yield and in addition to that positioning of quantum emitters is not deterministic \cite {RiedrichMller2014}. Thus scaling to multiple quantum emitters coupled with photons scales the difficulty.

\begin{figure}[htp!]
\centering
\includegraphics[width=0.95\columnwidth]{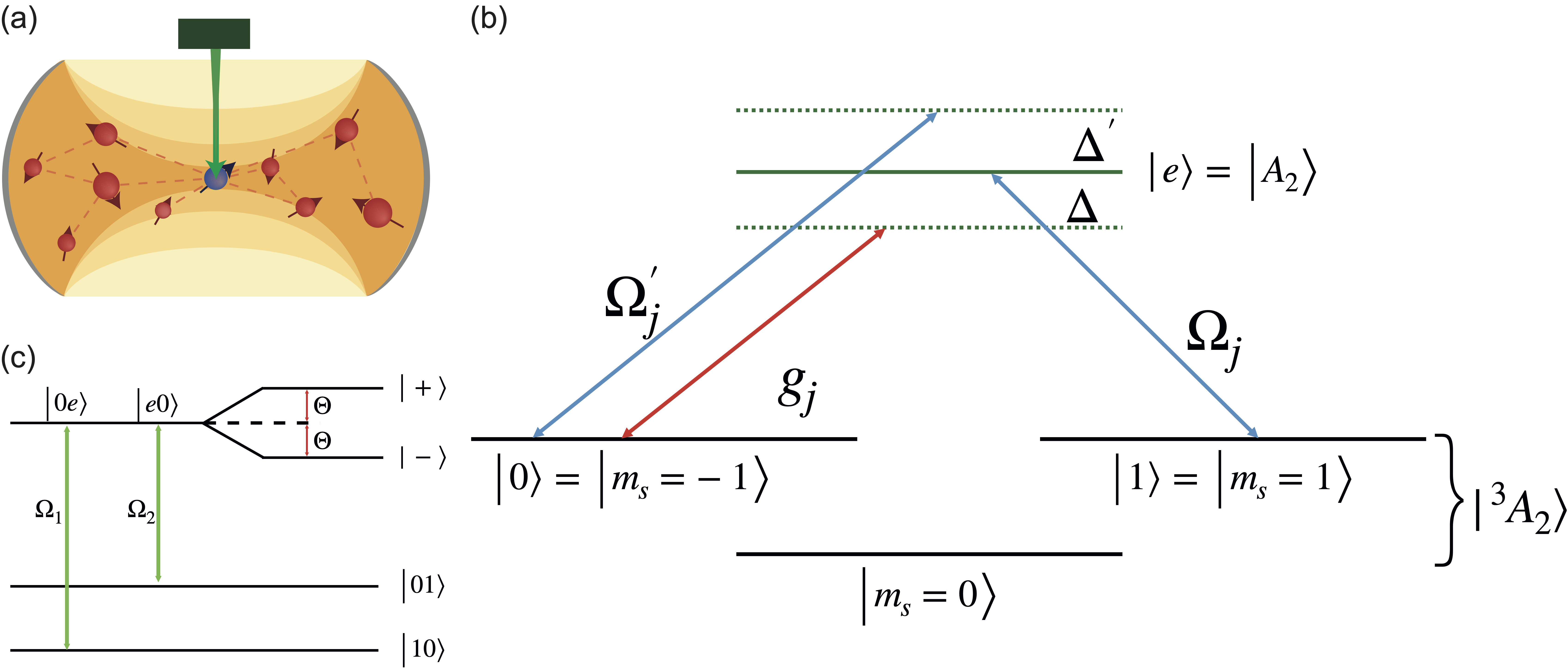}
\caption{\label{fig:CavityNV}(a) Schematic of NVs trapped in a cavity. The blue NV is chosen as the system spin and the rest of the NVs are considered as a part of the environment.(b)Energy level diagram of NV center with red (blue) line denoting cavity mode (laser field) coupling to the NV. (c) Raman transition between $NV_1$ and $NV_2$ in dressed state basis i.e $\ket{\pm} = (\ket{01}\pm\ket{10})/\sqrt{2}$ with coupling $\Theta = g_1g_2/\Delta$}
\end{figure}

Additionally, the interactions of individual NV center with cavity photon is measured with the Purcell enhancement of the spontaneous emission \cite{Zhang2018} and hybridization of the electronic state \cite{Togan2010}. However, for multiple emitters interacting with photons, although dipole-dipole interaction between the emitters is negligible, emitter-emitter interaction will exist through photon mediation. This will give rise to entanglement \cite{Angerer2018,Kubo2012}. Thus for many quantum emitters interacting with cavity photons, characterization using previous methods becomes nontrivial. No such direct methods exist, and closest methods that can be extended for characterization are those which are based on Double Electron-Electron Resonance (DEER) like techniques \cite{cooper2018spectral,belthangady2013dressed}. \add{Similar techniques \cite{laraoui2013high,ma2016proposal} which take advantage of 2D spectroscopy are present but the difference is that they have to utilize environment degrees of freedom which PrePSy doesn’t have to.} For such type of setups, we demonstrate  that PrePSy can quickly provide valuable insights.

As most of the system-environment properties are encoded in the correlations, we present PrePSy as an auxiliary characterizing technique for such systems where coupling between entities is sufficiently stronger than collective decoherence effects. With PrePSy, we demonstrate for multiple NVs trapped within a cavity the ability to read off parameters like coupling constant, the quantity of NVs coupled to the cavity.  We choose a single NV to be the system and the cavity, along with other NVs to be the environment, as shown in Figure \ref{fig:CavityNV}.a.

\subsection{Hamiltonian Dynamics}

The NV has an $S=1$ electronic ground state labeled as $\ket{^3A_2} = \ket{E_0}\otimes\ket{m_s = 0, \pm1}$ where $\ket{E_0}$ is the orbital angular momentum state and $\ket{m_s = 0, \pm1}$ are the spin angular momentum states. The optical transition between the ground and excited state manifold is spin preserving but changes the orbital angular momentum. The spin-photon coupling can be reduced to an effective pairwise photon mediated Jaynes-Cummings model with the help of laser-induced Raman transition between two centers via the exchange of virtual cavity photons \cite {Li2011}. 

To show this consider that the cavity mode in optical regime dispersively couples to the $j^{th}$ NV center's transition between the excited state ($\ket{e} = \ket{A_2} \coloneqq 1/\sqrt{2}(\ket{E_-}\ket{m_s = 1} + \ket{E_+}\ket{m_s = -1} )$ and a states in the ground state manifold ($\ket{0} \coloneqq \ket{E_0}\ket{m_s = -1}$), with coupling constant $g_j$ and detuning $\Delta$ as shown in Figure \ref{fig:CavityNV}.b. The selectivity of the states is because of the spin selectivity nature of the optical transition and this particular $\Lambda$-type transition was recently used for spin-photon interaction \cite {Togan2010}. Under the condition $\Delta \gg g_j$, the cavity operators can be adiabatically eliminated. To perform adiabatic elimination, time-averaging of the Hamiltonian is performed given that the cavity is detuned \cite{james2007effective}. The corresponding Hamiltonian is reduced to
\begin{equation}
\label{eqn:final}
 H_{\text{adiab}} = \sum_{\substack{i,j \\ i\neq j}}\frac{g_i*g_j}{\Delta}\ket{e_i 0_j}\bra{0_i e_j} + H.C.
\end{equation}

In addition to cavity interaction, a largely detuned $\sigma_+$ laser is coupled to the same transition with Rabi frequency $\Omega^{'}_j$ and detuning $\Delta^{'}$ ($\Delta \gg \Omega^{'}_j$). The purpose of this laser transition is to eliminate the stark shift term of the state $\ket{0}$ generated by the vacuum state of the cavity. 

Another adiabatic elimination of the excited state manifold of NVs is possible if a $\sigma_-$ laser is coupled to
$\ket{e} \leftrightarrow \ket{1}$ (where $\ket{1} = \ket{E_0}\ket{m_s = +1}$ and Rabi frequency is $\Omega_j$) to create a two photon Raman process Fig.\ref{fig:CavityNV}.C. Thus the effective Hamiltonian in the subspace of $\{\ket{10},(\ket{e0}+\ket{0e})/\sqrt{2}, (\ket{e0}-\ket{0e})/\sqrt{2},\ket{01}\}$ is 
\begin{equation}
    \hat{H}_{\text{eff}} = \sum_{\substack{i,j \\ i\neq j}} \xi_{i,j} \ket{1_i0_j}\bra{0_i1_j} + H.C.,
    \label{eqn:final2}
\end{equation}
where $\xi_{i,j} = \Omega_i\Omega_j/\Theta_{ij}$ and $\Theta_{ij} = g_ig_j/\Delta$. The adiabatic elimination is valid only when the effective NV - NV coupling is much larger than the laser coupling ( $\Theta_{ij} \gg \Omega_i , \Omega_j$). The derivation of the effective Hamiltonian as shown in Eqn. \ref{eqn:final2} which is twice adiabatically approximated is derived in section 4 of Supplementary.

If effective pairwise coupling is uniform i.e. $\xi_{i,j} = \xi\text{ }\forall\text{ }i,j$ then Hamiltonian for the group of two level systems can be written in terms of collective angular momentum operators ($J_x,J_y,J_z$) as 
\begin{equation}
    \hat{H}_{\text{eff}} = 4\xi\left(\vec{J}^{\,2} - J_z^2\right) - 2\xi N \cdot \mathbb{I},
    \label{eqn:apprHam}
\end{equation}
where N is the number of NVs in the cavity. The Hamiltonian derived is a type of non-linear rotator of spins and is called a one-axis twisting Hamiltonian. We simulate the density matrix evolution given in Eqn. \ref{eqn:lindblad} where $\hat{H}$ is given by Eqn. \ref{eqn:final}. 
\begin{align}
\label{eqn:lindblad}
    \begin{split}
        \dot{\rho}  = &-\iota\left[\hat{H},\rho\right] + \sum_{i}\gamma_{10}^{i}\left(2\hat{\sigma}^i_{01}\rho\hat{\sigma}^i_{10} - \left\{\hat{\sigma}^i_{01}\hat{\sigma}^i_{10},\rho\right\}\right) \\
        & + \sum_{i}\sum_{j = 0,1}\gamma_{ej}^{i}\left(2\hat{\sigma}^i_{je}\rho\hat{\sigma}^i_{ej} - \left\{\hat{\sigma}^i_{ej}\hat{\sigma}^i_{je},\rho\right\}\right),    
    \end{split}
\end{align}
The Hilbert space spanned by $\rho$ includes the system's (single NV) and the environment's (other NVs) Hilbert spaces. The system-environment is  weakly coupled to a super environment (bath). The decoherence processes due to this bath considered for the Lindblad master equation are spontaneous emission and spin-lattice de-excitation. Here, $\sigma_{\alpha \beta} = \ketbra{\alpha_i}{\beta_i}$, $\kappa$ is the cavity decay rate, $\gamma_{10}^{i}$ is the spin-lattice de-excitation rate and $\gamma^i_{ej}$ is the spontaneous emission from the excited state of NV.

\subsection{PrePSy on Cavity QED with NVs}
The NV-cavity setup weakly coupled to a super environment is initialized to Gibb's state at room temperature where the initial correlation between system NV and other NVs is primarily dependent on the interaction strength and decoherence rate. For the NV-Cavity system, the coupling strength of the interaction Hamiltonian can be few orders higher than the cavity decay rate\cite {Louyer2005,Spillane2005}. Hence the thermal state will have significant correlations between the system NV and other NVs in the cavity generated by photon-mediated coupling. The choice of the initial thermal state also simplifies state preparation.

\begin{figure}[htp!]
\centering
\includegraphics[width=0.85\columnwidth]{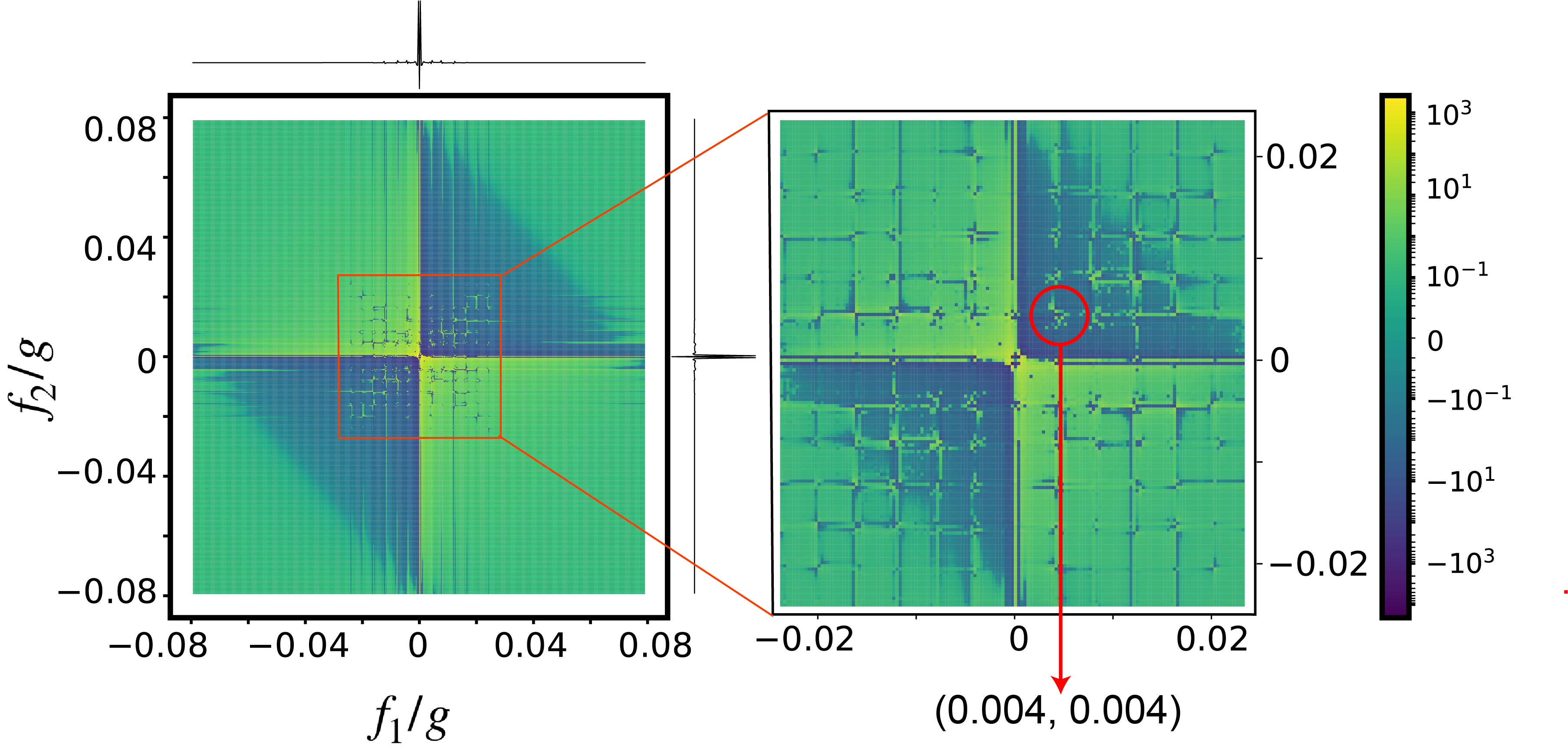}
\caption{\label{fig:6spins} PrePSy applied on a single NV placed in a cavity with an environment of five other NVs. Parameters are chosen as $\Delta = 10g$ and $\Omega = 0.01g$, $\xi = 0.001g$ }
\end{figure}

For applying PrePSy, conditional preparation is performed with for two variations of the projection operator $\ketbra{+x}{+x}$ and $\ketbra{-x}{-x}$ defined in the Hilbert subspace of $\ket{0} , \ket{1}$ and the system is then prepared to $\ket{+x}$. The $\pi/2$ rotation is generated by $\hat{\sigma}_y$ in the same Hilbert subspace. The final measurement is the population of the system NV in the state $\ket{+x}$.  Results for PrePSy's simulation for an open quantum system dynamics of 6 NV centers trapped in an optical cavity is shown in Figure \ref{fig:6spins} below where a single NV is considered as the system and the rest of the NVs are defined as the environment.

The result of PrePSy displays peaks placed uniformly in the frequency domain. Similar to standard 2D coherent spectroscopy, the position of the peak corresponds to the gap between the energy levels. Consider Hamiltonian for six spins given by Eqn. \ref{eqn:apprHam}, where the energy separation between two eigen-states ($\ket{j_h,m_h}$, $\ket{j_k,m_k}$) in term of quantum numbers $(j,m)$ is
\begin{equation}
    \label{eqn:gapformula}
    E_{h,k} = 4 \xi \left( \left(j_h-j_k\right) \left(j_h+j_k+1\right)+ \left(m_h^2-m_k^2\right)\right)
\end{equation}
For all possible combination of $(h,k)$ a peak will be observed in the 2D map. A peak at $(x,y)$ corresponds to energy transition $x$ during $t_1$ and energy transition $y$ during $t_2$ respectively. For example the smallest energy gap according to Eqn. \ref{eqn:gapformula} is between $\ket{0,0}$ and $\ket{1,1}$ which is $4\xi = 0.004$ and the peak closest to the origin in the figure at $(0.004,0.004)$. Thus from PrePSy pairwise coupling for simple systems can be easily calculated. At the positions predicted by the Eqn. \ref{eqn:gapformula}, multiple smaller peaks are visible rather than a single peak because of the adiabatic approximation. The simulation is performed for Eqn. \ref{eqn:final}, which does not assume adiabatic elimination of the excited state $\ket{e}$; however, the Eqn. \ref{eqn:gapformula} does.

Since there is decoherence in the form of spontaneous emission and cavity decay, higher energy levels eventually are depopulated, thus all peaks will not appear in the image and the higher frequency peaks are generally less visible.

The total number of peaks visible corresponds total number of transitions possible. For 6 spins there are $10$ distinct eigenenergy levels and $12$ distinct energies of the transition. However Figure \ref{fig:6spins} displays approximately $6$ distinct energies which indicates the presence of decoherence and is equivalent to $4$ spins interacting without decoherence. From this point of view, in the presence of noisy environment $6$ NVs in the cavity are effectively just $4$ NVs in the cavity without any decoherence.

\section{Conclusions \& Discussion}
PrePSy is capable of witnessing specific correlations, as demonstrated with the help of a toy model in Section 3. In addition to this, we show that PrePSy can also measure specific correlations because of the linear dependence of initial correlations on the signal measured.  In Section 4, we illustrate the potential of PrePSy to derive information related to the system-environment dynamics from the initial correlation using a cavity QED setup. More specifically, PrePSy measures the coupling constant of a spin-photon interaction between NVs and photons of an optical cavity and estimates the effective number of NVs interacting with the cavity in the presence of decoherence.

\begin{figure}[htp!]
\centering
\includegraphics[width=0.85\columnwidth]{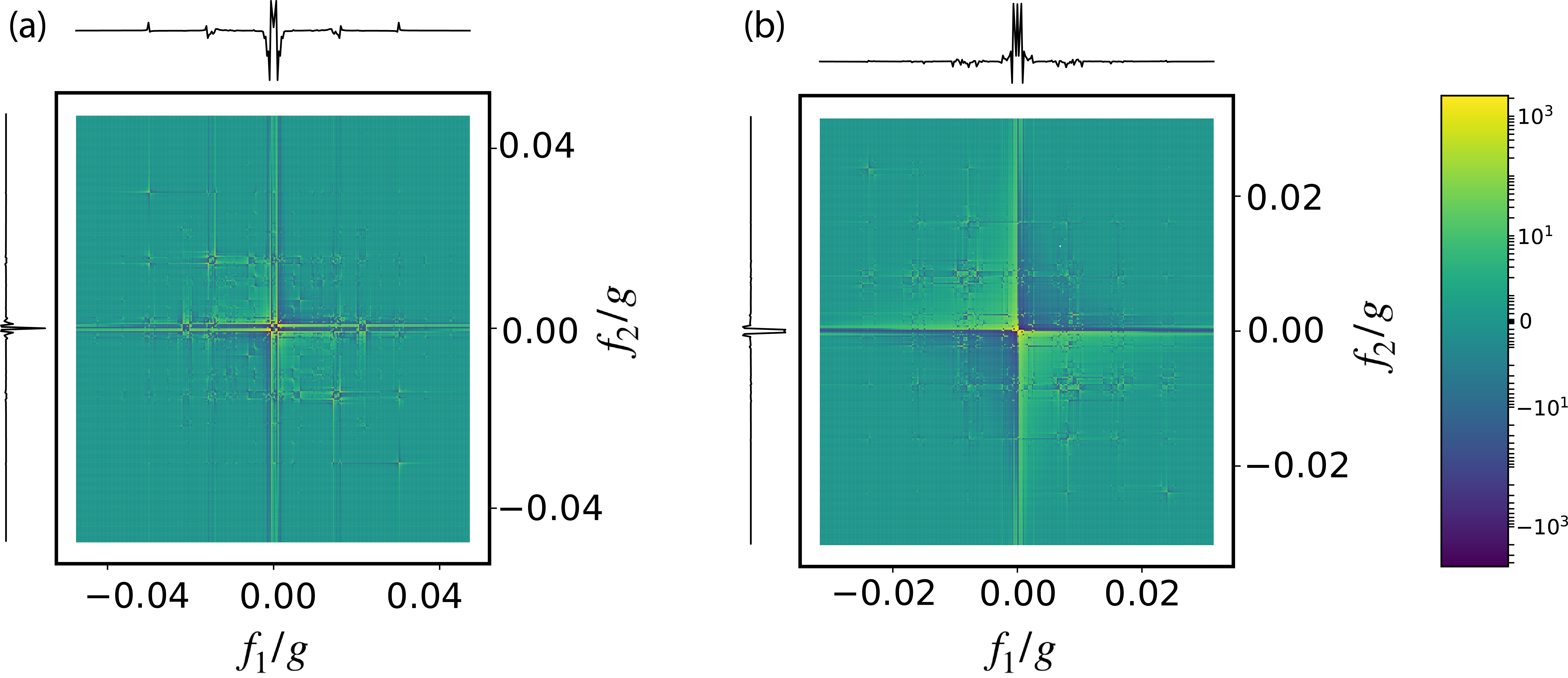}
\caption{\label{fig:3vs4spins}(a) PrePSy applied on a single NV which is part of a closed chain of 3 other NVs placed in a cavity. The Hamiltonian is $H_{\text{eff}} =\sum_{i=1}^{4}\ketbra{1_i0_{i+1}}{0_i1_{i+1}} + H.C.$ where $\ket{a_5} = \ket{a_0}$ for $a = 0,1$. The number of major peaks visible with 1D spectroscopy is 6, shown above the 2D image (ignoring the low frequency peaks). In 2D spectroscopy approximately 7 major peaks in the $2^{nd}$ and $4^{th}$ quadrant each. (b) PrePSy applied on a single NV which is in the cavity with two other NVs interacting in a pairwise manner. The Hamiltonian is $H_{\text{eff}} =\sum_{i,j; i\neq j}^{3}\ketbra{1_i0_j}{0_i1_j} + H.C.$. The number of major peaks visible with 1D spectroscopy is 6, shown above the 2D image (ignoring the low frequency peaks). In 2D spectroscopy approximately 9 major peaks in all quadrants.} 
\end{figure}

Experimentally deploying PrePSy is feasible for a vast array of practically relevant physical systems since PrePSy assumes little about the system-environment state and is motivated for generic quantum systems. For color center - cavity QED  systems, PrePSy is achievable with current experimental techniques \cite{wan2020large, bhaskar2017quantum, englund2010deterministic} where experimentally spin-photon coupling has been demonstrated. The only requirement for applying PrePSy is a clear demarcation of system and environment. The system should be such that it is amenable to measurement, projection and rotation. Moreover, pulse sequences applied to the system should not affect the environment beyond producing conditional post-measurement states. This demarcation can restrict the physical systems PrePSy is applicable to. An example in NMR is the inability to separate out a single molecule as the system and the rest as environment because it works on ensemble statistics.

When experimentally performing PrePSy, 2D spectroscopy is preferred as it aids in quantifying the type of interaction Hamiltonian. As population transfer channels vary with different interaction Hamiltonian, so will the 2D image obtained. However, for 1D spectroscopy, two different Hamiltonians can give results that are not differentiable. For example, if 1D spectroscopy is applied to 3 NVs with pairwise interactions and a closed chain of 4 NVs, the same number of peaks will be visible for both the cases. Hence 2D spectroscopy differentiates between these two cases. The result of PrePSy for both these cases, as shown in Figure \ref{fig:3vs4spins}, is strikingly different. 

As our method scales favorably with the system size, it can be conveniently applied on NVs with any other form of ancilla \cite{rabl2009strong,li2016hybrid,meesala2016enhanced}. We hope this will find applications in characterizing complex quantum systems that have initial or intermediate correlations.

\section*{Acknowledgements}
KS acknowledges support from IITB-IRCC Seed grant number 17IRCCSG009, DST QUEST - Project 4, Q-58 and DST Inspire Faculty Fellowship - DST/INSPIRE/04/2016/002284. SV acknowledges support from an IITB-IRCC grant number 16IRCCSG019, the National Research Foundation, Prime Ministers Office, Singapore under its Competitive Research Programme (CRP Award No. NRF-CRP14-2014-02), DST-SERB Early Career Research Award (ECR/2018/000957) and  DST-QUESTgrant number DST/ICPS/QuST/Theme-4/2019.

\section*{Author Contributions}
S.V. and P.J. designed the Prepare Probe Spectroscopy sequence and simulated the toy model. K.S. and P.J worked on the application of the PrePSy scheme on Nitrogen Vacancy centers coupled to an optical cavity. K.S. and P.J. numerically demonstrated that PrePSy can be used to calculate system parameters for NVs coupled to cavity. P.J. analyzed the data and prepared the manuscript in discussion with S.V. and K.S. All the authors have read and approved the final version of the manuscript. 

\section*{Competing Interests}
The authors declare no competing interests. 

\section*{Appendix}
\appendix
\section{Conditional Preparation as a mean to probe specific correlations}
\label{appendix:condprep}
Post-conditional preparation the system-environment state is $\ketbra{0}{0}\otimes\left\{\text{Tr}_s\left[\left(\ketbra{m}{m}\otimes\mathbb{I}_E\right)\chi\right]+\tau\right\}$ as defined in Eqn. \ref{eqn:postcp}. The correlation matrix $\chi$ can be written in form of system states ($\ket{i_s}$) and environment states ($\ket{i_e}$) as 
\begin{equation}
    \chi = \sum_{\substack{i_s,j_s \\ i_e, j_e}} \chi_{\substack{i_s,j_s \\ i_e, j_e}} \ketbra{i_s i_e}{j_s j_e}.
\end{equation}
Substituting back in $P \coloneqq \left(\ketbra{m}{m}\otimes \mathbb{I}_E\right)\cdot\chi$ gives
\begin{equation}
P = \sum_{\substack{i_s,j_s \\ i_e, j_e}}\chi_{\substack{i_s,j_s \\ i_e, j_e}}\left(\ketbra{m}{m}\otimes \mathbb{I}_E\right)\left(\ketbra{i_si_e}{j_sj_e}\right),
\label{eqn:rho0eexp}
\end{equation}
where $P$ is post entanglement breaking channel part of environment which remembers the initial correlation. Partial tracing over the system gives
\begin{equation}
    \text{Tr}_s(P) = \sum_{\substack{i_s,j_s \\ i_e, j_e}}\chi_{\substack{i_s,j_s \\ i_e, j_e}}\braket{j_s}{m}\braket{m}{i_s}\ketbra{i_e}{j_e}.
\end{equation}
\nocite{*}
Thus for $\chi_{i,j}$ (where $i = i_s N_s + i_e$, $j = j_s N_s + j_e$ and $N_s$ is the Hilbert space dimension of the system) to be retained after conditional preparation $\braket{m}{i_s}$ and $\braket{j_s}{m}$ should not be zero. Hence the projection operator $\ketbra{m}{m}$ should not be orthogonal to $\ket{i_s}$ and $\ket{j_s}$.

\section{Conditions for the signal to not register a finite \texorpdfstring{$\chi_{i,j}$}{chi-ij}}

\label{appendix:Hamil}
For $\chi_{i,j}$ independent signal, the derivative of signal with respect to $\chi_{i,j}$ should be zero. If the final measurement operator is $\ketbra{n}{n}$, the measured signal is
\begin{equation}
    N(t_1,t_2) = Tr\left[\left(\ketbra{n}{n}\otimes \mathbb{I}_E\right)\rho^{SE}(t_1,t_2)\right].
\end{equation}

$\ketbra{n}{n}$ and $\rho_{SE}(t)$ can be written in their element-wise format giving $N(t_1,t_2)$ as
\begin{equation}
= \text{Tr}\left(\sum_{k,l} n_{kl}\ketbra{k}{l}\otimes\mathbb{I}_E\cdot\sum_{\substack{i_s,j_s \\ i_e, j_e}}\rho^{SE}_{\substack{i_sj_s \\ i_e j_e}}(t_1,t_2)\ketbra{i_s i_e}{j_s j_e}\right).
\end{equation}
On simplification gives \add{$N = \sum_{k} n_{kk} \sum_{i_e}\rho^{SE}_{kki_ei_e}(t_1,t_2)$} where, $\rho^{SE}(t_1,t_2)$ can be written using system-environment unitary evolution operators as 
\begin{equation}
  \rho^{SE}(t_1,t_2)=  e^{-\iota H t_2}e^{-\iota \frac{\pi}{2} A}e^{-\iota H t_1}\rho_{SE}(0)e^{\iota H t_1}e^{\iota \frac{\pi}{2} A}e^{\iota H t_2}.
\end{equation}
Simply can be written as $\mathcal{U}\cdot\rho_{SE}(0)\cdot\mathcal{U}^{\dagger}$ for now, where $\mathcal{U}$ describes the complete unitary evolution, $H$ is the system-environment total Hamiltonian and $\hat{A}$ is a rotation operator. Here $\rho^{SE}(0)$ is given by Eqn. \ref{eqn:rho0}, where the element expansion is given Eqn. \ref{eqn:rho0eexp}. Substituting back, differentiating with $\chi_{i^{'}_s, j^{'}_s, i^{'}_e, j^{'}_e}$ and on simplifying,\begin{equation}
    \frac{\partial N}{\partial \chi_{\substack{i^{'}_s,j^{'}_s \\i^{'}_e, j^{'}_e}}} = \sum_{k,i_e}n_{kk}m^{'}_{j_s}m^{'}_{i_s}\left[\mathcal{U}\cdot\ketbra{0,i^{'}_e}{0,j^{'}_e}\cdot\mathcal{U}^{\dagger}\right]_{\substack{k,k,\\i_e,i_e}},
    \label{eqn:fulleqn}
\end{equation}
where $m^{'}_{i_s}$ ($m^{'}_{j_s}$) $ = \braket{m}{i^{'}_s}$ ($\braket{m}{j^{'}_s}$). If the projection operator $\ketbra{m}{m}$ is not orthogonal to $\ket{i^{'}_s}$ and $\ket{j^{'}_s}$, then $m^{'}_{i_s}$,$m^{'}_{j_s}$ are non zero. For signal to be independent Eqn. \ref{eqn:fulleqn} should be zero. This will lead to 
\begin{equation}
    \sum_{k,i_e}n_{kk}\left[\mathcal{U}\cdot\ketbra{0,i^{'}_e}{0,j^{'}_e}\cdot\mathcal{U}^{\dagger}\right]_{k,k,i_e,i_e} = 0.
\end{equation}
Thus for $\chi_{i^{'},j^{'}}$ independence, there should be no population change in $\ketbra{n}{n}$ due to $\ketbra{0,i^{'}_e}{0,j^{'}_e}$ under PrePSy. Hence there should exist seperate Hilbert subspaces consisting of $\ketbra{0,i^{'}_e}{0,j^{'}_e}$ and $\ketbra{n}{n}$ respectively under action of Hamiltonian (H) and $\pi/2$ rotation due to $\hat{A}$.

\section{Characterizing initial correlation using only one-dimensional Spectroscopy}
\label{Appendix:1dprespsy}
One-dimensional spectroscopy for PrePSy is convenient for quickly observing and measuring the initial correlations due to the reduced experimental effort. Only having single-dimensional spectroscopy implies the computational complexity is reduced, and hence parameters of more complex systems can be obtained from fitting the data. The drawback of one-dimensional spectroscopy is the ambiguity in accurately calculating the correlations and Hamiltonian as the output of PrePSy, which is no more unique and is shown in conclusion.

\begin{figure}[htb]
\centering
\includegraphics[width=1\columnwidth]{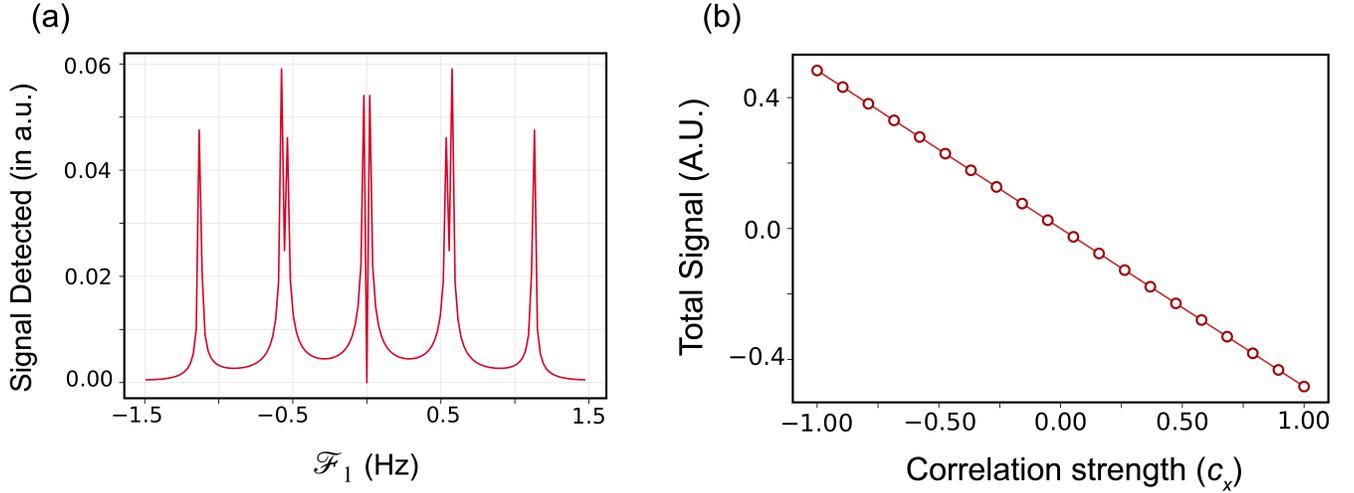}
\caption{\label{fig:1dprepsy} One dimensional PrePSy applied on the toy model. (a) 1D Date of PrePSy with parameters as $c_x = 1$, $\lambda^{xx} = 4Hz$, $\lambda^{yy} = 3Hz$ and $\lambda^{zz} = 3.5Hz$. (b) Plot of the variation in total signal strngth vs the initial correlation $c_x$.} 
\end{figure}

For the toy model described in Section A.a, one-dimensional PrePSy was carried with the same parameters. The results obtained are shown in Figure \ref{fig:1dprepsy}, where the initial projection operators ($\ketbra{x}{x}$ and $\ketbra{-x}{-x}$)  and the measurement operator ($\ketbra{x}{x}$) are kept the same.  As the configuration of PrePSy is identical, the positions of the peaks will not change. However, the shapes of the peaks will change as the spectroscopy process is different. The presence of peaks indicates the presence of correlations.

Similar to the two-dimensional case, the total correlation varies linearly with the amount of initial correlation present between the system and the environment, as seen in Figure \ref{fig:1dprepsy}. Hence the method to measure correlation remains the same as discussed in Section 3.B

\section{Adiabatic Elimination of the cavity}
\label{Appendix:AdiabDerive}
The method to obtain the effective Hamiltonian by adiabatic elimination under dispersive coupling regime is based on time-averaging of the Hamiltonian \cite{james2007effective}. A perturbative Hamiltonian with a harmonic time dependence is of the form
\begin{equation}
    \hat{H}_{I}(t)=\sum_{n=1}^{N} \hat{h}_{n} \exp \left(-i \omega_{n} t\right)+\hat{h}_{n}^{\dagger} \exp \left(i \omega_{n} t\right),
\end{equation}
where N is the total number of different harmonic terms making up the interaction Hamiltonian and $\hat{h}_n$ are the operators defining the interactions. The effective Hamiltonian obtained by letting the fast evolving terms average to zero, is
\begin{equation}
\label{eqn:AdiabElim}
    \hat{H}_{e f f}(t)=\sum_{m, n=1}^{N} \frac{1}{\overline{\hbar \omega}_{m n}}\left[\hat{h}_{m}^{\dagger}, \hat{h}_{n}\right] \exp \left(i\left[\omega_{m}-\omega_{n}\right] t\right),
\end{equation}
where $\overline{\omega}_{mn}$ is the harmonic average of $\omega_m$ and $\omega_n$, viz.,
\begin{equation}
    \frac{1}{\bar{\omega}_{m n}}=\frac{1}{2}\left(\frac{1}{\omega_{m}}+\frac{1}{\omega_{n}}\right).
\end{equation}

Thus using this method the lab frame interaction Hamiltonian of just ``N'' NVs interacting with the cavity field is as follows
\begin{equation}
    \hat{H} = \sum_i^N \left[g_i\left(a^{\dagger}\ketbra{0_i}{e_i}e^{i\Delta t} + a\ketbra{e_i}{0_i}e^{-i\Delta t}\right)\right].
\end{equation}
Applying Eqn. \ref{eqn:AdiabElim} on the above Hamiltonian to get the effective Hamiltonian as
\begin{align}
    \begin{split}
        \hat{H}_{\text{eff}} &= \sum_{\substack{i,j \\ i\neq j}}^N \frac{\hbar g_i g_j}{\Delta}\ketbra{0_ie_j}{e_i0_j} + \ketbra{e_i0_j}{0_ie_j} \\
        &+ \sum_{i=1}^{N}\frac{\hbar g^2}{\Delta}\left(\ketbra{e_i}{e_i} - \ketbra{0_i}{0_i}\right)\left(2\hat{a}^{\dagger}\hat{a} - 1\right).
    \end{split}
\end{align}
The second term is the quantum Stark term, which is eliminated by firstly using a laser coupled to the same transition and secondly by assuming that the cavity is initially in the vacuum state. Additionally, a laser with zero detuning and Rabi frequency $\Omega$ is coupled to $\ket{1_i}\leftrightarrow \ket{e_i}$ transition to generate Raman-like transition. Thus the Hamiltonian is 
\begin{align}
\begin{split}
     H_{\text{adiab}} &=  \sum_{\substack{i,j \\ i\neq j}}^N \left[\frac{g_ig_j}{\Delta}\ketbra{e_i0_j}{0_ie_j}\right] 
     + \sum_i^N \left[\Omega_i\left(\ketbra{e_i}{1_i} + \ketbra{1_i}{e_i}\right)\right].
\end{split}
\end{align}

Applying a change of basis $\ket{+_{ij}} = (\ket{e_i0_j}+\ket{0_ie_j})/\sqrt{2},\text{ }\ket{-_{ij}} = (\ket{e_i0_j}-\ket{0_ie_j})/\sqrt{2}$. The Hamiltonian in the new basis can be seperated into two non-interacting subspaces. Choosing the subspace $\{\ket{10},\ket{+},\ket{-},\ket{01}\}$ and assuming the initial state to be in the subspace, the other subspace can be ignored. The Hamiltonian in this subspace is 
\begin{align}
\begin{split}
    H &= \sum_{\substack{i,j \\ i\neq j}}^N\ketbra{+_{ij}}{+_{ij}}\left(\Theta_{i,j}\right) - \ketbra{-_{ij}}{-_{ij}}\left(\Theta_{i,j}\right) \\
     &+ \sum_i^N \Lambda_i\ketbra{+}{10} + \Lambda_i\ketbra{-}{10}+ H.C.,
\end{split}
\end{align}
where $\Theta_{i,j} = g_ig_j/\Delta$ and $\Lambda_i = \Omega_i/\sqrt{2}$. The schematic diagram of the interaction is given by Figure \ref{fig:CavityNV}.C. Applying adiabatic elimination given by Eqn. \ref{eqn:AdiabElim} again, the effective Hamiltonian obtained is given by Eqn. \ref{eqn:final2}.

\end{document}